\documentclass[conference]{IEEEtran}
\IEEEoverridecommandlockouts

\usepackage{booktabs} 
\usepackage{hyperref} 

\usepackage{cite}
\usepackage{amsmath,amssymb,amsfonts}
\usepackage{algorithmic}
\usepackage{graphicx}
\usepackage{textcomp}
\usepackage{xcolor}
\def\BibTeX{{\rm B\kern-.05em{\sc i\kern-.025em b}\kern-.08em
    T\kern-.1667em\lower.7ex\hbox{E}\kern-.125emX}}
\begin{document}

\title{Quantifying the Echo Chamber Effect: \\ An Embedding Distance-based Approach}

\author{Faisal Alatawi, Paras Sheth, Huan Liu \\
Arizona State University \\
\{faalataw,psheth5,huanliu\}@asu.edu
}


\maketitle

\begin{abstract}
The rise of social media platforms has facilitated the formation of echo chambers, which are online spaces where users predominantly encounter viewpoints that reinforce their existing beliefs while excluding dissenting perspectives. This phenomenon significantly hinders information dissemination across communities and fuels societal polarization. Therefore, it is crucial to develop methods for quantifying echo chambers. In this paper, we present the Echo Chamber Score (ECS), a novel metric that assesses the cohesion and separation of user communities by measuring distances between users in the embedding space. In contrast to existing approaches, ECS  is able to function without labels for user ideologies and makes no assumptions about the structure of the interaction graph.
To facilitate measuring distances between users, we propose EchoGAE, a self-supervised graph autoencoder-based user embedding model that leverages users' posts and the interaction graph to embed them in a manner that reflects their ideological similarity. To assess the effectiveness of ECS, we use a Twitter dataset consisting of four topics - two polarizing and two non-polarizing. Our results showcase ECS's effectiveness as a tool for quantifying echo chambers and shedding light on the dynamics of online discourse.
\end{abstract}


\section{Introduction}

In the age of digital communication, social media platforms have revolutionized the way we disseminate and consume information. Nevertheless, this evolution has brought about notable challenges, particularly the emergence of echo chambers and polarization~\cite{Cinelli2021Echo,Terren2021Echo,Alatawi2021Survey}. These phenomena are often characterized by high levels of controversy between members of different groups and homogeneity among members of the same group~\cite{Villa2021Echo}. This reinforces pre-existing beliefs~\cite{Cinelli2021Echo}, discourages critical thinking~\cite{Rhodes2022Filter}, promotes the spread of misinformation~\cite{DelVicario2016Spreading,DiazRuiz2023Disinformation}, and leads to societal divisions. Hence, it is crucial to devise methods for measuring the extent and impact of echo chambers on social media. By quantifying them, we can better understand these phenomena and, consequently, devise strategies to mitigate echo chamber effects and foster more balanced and nuanced discussions. Ultimately, this could contribute to a better informed, open-minded, and empathetic society. Such efforts are particularly crucial in today's world, where topics such as politics, health, economics, and environmental issues, which are susceptible to echo chambers~\cite{Baumann2020Modeling,Villa2021Echo}, have far-reaching implications for society.

Echo chambers are contingent on two dynamics: the interaction among users and the individual ideological leanings of these users. Numerous measures and metrics have been developed to leverage these dynamics, either separately or in conjunction. One such method, is to leverage the interactions graph to compute graph-specific metrics such as modularity~\cite{Wolfowicz2023Examining}, or resort to other techniques like random walkers~\cite{Garimella2018Quantifying}. However, utilizing the graph introduces a difficulty, as a graph may exhibit modularity without necessarily being polarized or containing an echo chamber~\cite{Guerra2013Measure,Wolfowicz2023Examining}. An alternate approach involves assessing the ideological disparity between users and their adjacent nodes within the graph, investigating correlations between a user's ideology and that of their neighbors~\cite{Cota2019Quantifying,Cinelli2021Echo}, or observing ideological deviations from the center of an opinion scale after deploying opinion-spreading models~\cite{Matakos2017Measuring,Musco2018Minimizing}. These methodologies, although insightful, are fraught with challenges. Labeling users to ascertain their ideologies or opinions is a laborious task that is susceptible to errors. Similarly, semi-supervised methods that depend on weak labels also present their own unique set of complications. 

In response to these issues, we introduce the Echo Chamber Score (ECS) a metric that captures the essence of the echo chamber concepts by focusing on the dynamic interactions both within and across different user communities. The crux of our approach is to gauge the similarity of users with their respective communities (i.e., cohesion) and across different communities (i.e., separation). Here, an interaction graph can be characterized as exhibiting an echo chamber-like structure if it exhibits a low average distance between users of a single community (i.e., high cohesion) and a high average distance between users across different communities (i.e., high separation). This strategy of using the distance allows us to bypass reliance on potentially incidental graph structures and eliminates the need to split the graph into two separate communities, an action that erroneously assumes inherent polarization. Further, our method uses similarity in the embedding space as a proxy for ideological distance, thereby circumventing the arduous and error-prone task of detecting individual users' ideologies. 

To facilitate the measurement of ideological distance, we propose EchoGAE, a self-supervised Graph Auto-Encoder~\cite{Kipf2016Variational} (GAE) based user embedding model. EchoGAE is designed to capture the ideological similarities among users through their interactions and shared posts, operating on two core principles: homophily~\cite{McPherson2001Birds}, where individuals associate and interact with those similar to themselves, and linguistic homophily~\cite{Yang2017Overcoming,Kovacs2020LanguageStyle}, the tendency of socially connected users to use language in similar ways. EchoGAE leverages homophilic interactions such as retweets, regarded as endorsements of similar ideologies~\cite{Garimella2018Quantifying,Cossard2020Falling}, along with the content of user posts. Both serve as inputs to capture and map these ideological similarities. The model architecture comprises an encoder that positions similar nodes closely together in the embedding space, and a decoder that uses users' embedding to reconstruct the graph structure in a self-supervised manner. Additionally, it utilizes Sentence-BERT~\cite{Reimers2019SentenceBERT}, a BERT-based language model, to embed tweets, thus reflecting their semantic similarities. By uniquely combining the interaction graph structure and linguistic information from user posts, EchoGAE generates representations that accurately reflect ideological similarities, establishing it as a robust tool for measuring the effects of echo chambers and polarization.

In this research, we evaluate the ability of the Echo Chamber Score (ECS) to measure echo chamber effects within homophilic social interaction networks. Our experiments are based on real-life Twitter datasets related to four topics: two polarizing and two non-polarizing. Our findings confirm that the ECS metric accurately identifies polarized interaction graphs and quantifies the echo chamber effect in a manner consistent with existing state-of-the-art methods. Furthermore, ECS proves successful in determining which communities within the interaction graph are more polarized, demonstrating its unique ability to rank communities based on their polarization. We also verify that EchoGAE's user embedding effectively reflects ideological distances between users, showcasing its capacity to detect user ideologies. To promote reproducibility and foster further development in this field, we make our datasets and code available to the public\footnote{\href{https://github.com/faalatawi/echo-chamber-score}{https://github.com/faalatawi/echo-chamber-score}}.

\section{Related Work}
Echo chambers and polarization measures can be divided into two main types: graph-based and ideology-based methods. \textbf{Graph-based methods} are based on the concept of a graph representing interactions between users on a given topic. These methods operate on the assumption that polarization can be observed within the graph itself. For instance, the modularity of a graph, which quantifies how well a graph can be divided into distinct communities, has been used to measure echo chambers~\cite{Wolfowicz2023Examining}. However, challenges arise from this approach, as modularity and other similar methods may not accurately represent echo chamber phenomena due to the possibility that non-polarized graphs can also exhibit high modularity~\cite{Guerra2013Measure,Wolfowicz2023Examining}.

To address these limitations, new methods have been developed that scrutinize the interactions between communities within a graph. These improved methods involve dividing the graph into two distinct communities and measuring polarization at the boundaries between them~\cite{Guerra2013Measure}. An alternative approach involves using the Random Walk Controversy~\cite{Garimella2018Quantifying} (RWC), a popular polarization method~\cite{Rashed2021EmbeddingsBased,Cossard2020Falling} that calculates the probability of a random walker starting at one community and ending at another. Nonetheless, these methods have their own drawbacks, such as the necessity of splitting the communities in the graph and making an inherent assumption that the graph is already polarized. This results in difficulties in measuring polarization that may not actually exist.

Our novel approach, the Echo Chamber Score (ECS), alleviates these issues. The ECS does not require the division of the graph into two communities and is capable of measuring the effects of echo chambers and polarization across any number of communities, making it a more flexible and accurate method for assessing polarization.

\textbf{Ideology-based methods} for measuring echo chambers and polarization take a different approach, focusing on a user's ideological leaning and the users they interact with. Two primary approaches exist within this category: (1) measuring the ideological distance between a user and their neighboring users in the graph, and (2) measuring the divergence from an ideological center after applying an opinion-spreading model.

In the first approach, the ideological leanings of all users are estimated and then compared to their neighboring users. The fundamental idea here is that an echo chamber is formed when users mostly interact with others who share similar opinions~\cite{Cota2019Quantifying, Cinelli2021Echo, Barbera2015Tweeting}. For instance, the ideology of users can be inferred from the hashtags they share or the content they post~\cite{Cota2019Quantifying, Cinelli2021Echo}. The polarization is then quantified by measuring the Pearson correlation between a user's ideological score and the average ideological score of their neighbors~\cite{Cota2019Quantifying, Cinelli2021Echo}.

In the second approach, opinion-spreading models such as the Friedkin-Johnsen or DeGroot opinion model are utilized \cite{DeGroot1974Reaching, Matakos2017Measuring, Musco2018Minimizing, Morales2015Measuring}. For instance, the Friedkin-Johnsen model operates by updating a node's opinion through repeatedly averaging the opinions of its neighbors until reaching equilibrium~\cite{Matakos2017Measuring, Musco2018Minimizing}. Polarization is then measured by how much opinions at equilibrium deviate from the average~\cite{Matakos2017Measuring, Musco2018Minimizing}. Alternatively, the DeGroot opinion model is used to construct a Polarization Index (PI) based on the probability density distribution of individuals' opinions \cite{Morales2015Measuring}. A bimodal distribution would suggest the existence of polarization, while a unimodal distribution would indicate its absence \cite{Morales2015Measuring}.

Both these ideology-based approaches have challenges, such as the laborious and error-prone task of estimating users' ideological leanings from their content or interactions. Therefore, we have opted instead for a model based on similarity in the embedding space as a proxy for ideology, eliminating the need for ideology estimation.

\section{Methodology}
This section presents our approach to quantifying echo chambers in online conversations. Our objective is to assess whether the discussion surrounding a given topic exhibits polarization and whether the communities formed by users can be characterized as echo chambers or comprise a diverse group of individuals with varying ideologies. To achieve this, we construct a graph $G = (V, E)$, where $V$ represents the set of social media users, and $E$ represents the edges denoting homophilic interactions, such as retweets. Additionally, we obtain a set of communities $\Omega$ from a community detection algorithm, where each community consists of a group of users. Our primary aim is to measure the level of polarization within the entire graph by computing the Echo Chamber Score (ECS) for each community. Consequently, this section presents our novel ECS metric for quantifying echo chambers. However, as ECS relies on user embedding, we begin by introducing our user embedding framework, EchoGAE, which enables the representation of users based on their ideological similarity.

\subsection{Embedding Social Media Users}
\begin{figure*}[]
    \centering
    \includegraphics[width=0.8\textwidth]{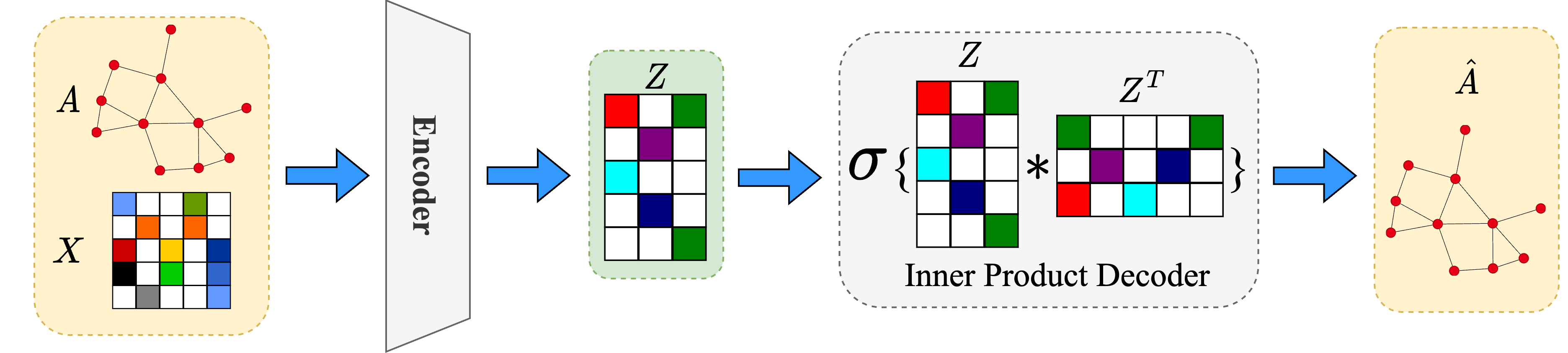}
    \caption{
    The EchoGAE Model comprises two primary components: an Encoder and a Decoder. The Encoder employs both user content embeddings ($\mathbf{X}$) and adjacency matrix ($\mathbf{A}$) to generate user embeddings ($\mathbf{Z}$). The Decoder then reconstructs the adjacency matrix ($\mathbf{\hat{A}}$) using the user representations.
    }
    \label{fig:EchoGAE}
\end{figure*}

The EchoGAE model (see figure~\ref{fig:EchoGAE}) is essential to our methodology for quantifying echo chambers in online conversations. Its purpose is to embed users in a way that reflects their ideological similarity, facilitating the calculation of the Echo Chamber Score (ECS). By placing ideologically similar users closer in the embedding space, EchoGAE enables the measurement of cohesion and separation of communities in the graphs, the two components of ECS, as we will explain later in the section.

EchoGAE is an adaptation of the Graph Auto-Encoder (GAE) model~\cite{Kipf2016Variational}, tailored for user embedding based on tweets and interactions. As a self-supervised model, EchoGAE eliminates the need for user ideological labeling. It employs two graph convolutional layers to encode the graph into a latent representation, which is subsequently decoded to reconstruct the graph structure. EchoGAE aims to minimize the binary cross-entropy between the real and reconstructed adjacency matrices.

The EchoGAE model consists of two main components: an Encoder and a Decoder. The Encoder takes both the tweets and the graph as input to create node embeddings, which serve as the user embeddings. The Encoder is divided into two parts. Firstly, the tweets component utilizes Sentence-BERT~\cite{Reimers2019SentenceBERT} to embed the user's tweets, and the average of these tweet embeddings is taken to form the content embeddings (In fig~\ref{fig:EchoGAE}, it's represented as the matrix $\mathbf{X}$). Secondly, the network component leverages the adjacency matrix ($\mathbf{A}$ in fig~\ref{fig:EchoGAE}) of the graph. Together, these components contribute to the creation of nodes embeddings (or users embeddings $\mathbf{Z} \in \mathbb{R}^{n \times d}$ where $n$ is the number of users in the graph and $d$ is the dimension of user embedding) that capture the information from both the users' content and their network interactions.

The Decoder performs an inner product operation~\cite{Kipf2016Variational} on the node representations ($\sigma(\mathbf{Z} * \mathbf{Z^T})$) obtained from the Encoder, resulting in a reconstructed adjacency matrix ($\mathbf{\hat{A}}$). Subsequently, the binary cross-entropy loss is used to train the model and ensure accurate graph reconstruction.

\subsection{Measuring the Echo Chamber Effect}

We introduce ECS (Echo Chamber Score), a measure for quantifying the echo chamber and polarization effects on social media. To measure the echo chamber effect using user embedding, we assess in-group cohesion and between-group separation \cite{Xiong2014Clustering}. We utilize the distance in the embedding space as a proxy for these factors, reflecting how closely related users within a community are (cohesion) and how distinct a community is from others (separation).

Let $Z \in \mathbb{R}^{n \times d}$ represent user embeddings, where $n$ is the number of users and $d$ is the embedding dimension. Additionally, let $\Omega = \{\omega_1, \omega_2, \dots, \omega_M\}$ denote the set of communities, where $\omega_i \subset V$ represents the $i^{th}$ community consisting of users. For a user $u \in \omega$, we compute the cohesion value ($\lambda_u$) as the average distance between $u$ and other users in the same community using Equation~\ref{eq:cohesion}.

\begin{equation}
    \lambda_u = \frac{1}{|\omega|} \sum_{\substack{v \in \omega \\ v \neq u}} dist(u, v)
    \label{eq:cohesion}
\end{equation}

Here, $|\omega|$ denotes the number of users in the community $\omega$, and $dist(u, v)$ represents the distance (e.g., Euclidean) between users $u$ and $v$ in the embedding space ($Z^{(u)}$ and $Z^{(v)}$ respectively). Similarly, we compute the separation value ($\Delta_u$) as the average distance between $u$ and the nearest community other than $\omega$ using Equation~\ref{eq:separation}.

\begin{equation}
    \Delta_u = \operatorname*{min}_{\substack{\omega \in \Omega \\ u \notin \omega}} \left [ \frac{1}{|\omega|} \sum_{v \in \omega} dist(u, v)      \right ]
    \label{eq:separation}
\end{equation}

To calculate the Echo Chamber Score (ECS) for a community $\omega = \{u_1, u_2, \dots, u_N\}$, we use a formula inspired by the silhouette score \cite{Rousseeuw1987Silhouettes} (in the appendix we show how to derive the ECS from the silhouette). Equation~\ref{eq:community_echo_chamber_score} produces a score between 0 and 1, with a higher score indicating a greater likelihood of an echo chamber effect within the community.

\begin{equation}
    ECS^{*}(\omega) = \frac{1}{|\omega|} \sum_{u \in \omega} \frac{max(\Delta_u, \lambda_u) + \Delta_u - \lambda_u}{ 2 * max(\Delta_u, \lambda_u)}
    \label{eq:community_echo_chamber_score}
\end{equation}

The Echo Chamber Score can be computed for the entire graph using Equation~\ref{eq:graph_echo_chamber_score}, where $\Omega$ represents the set of communities obtained from a community detection algorithm such as Louvain \cite{Blondel2008Fast} or Leiden \cite{Traag2019Louvain}.

\begin{equation}
    ECS(\Omega) = \frac{1}{|\Omega|} \sum_{\omega \in \Omega} ECS^{*}(\omega)
    \label{eq:graph_echo_chamber_score}
\end{equation}

The Echo Chamber Score (ECS) allows for comparison across different graphs representing various controversial topics. A higher ECS indicates a higher degree of echo chamber within a conversation. The components of ECS can provide additional insights, such as ranking communities based on their polarization, using Equation~\ref{eq:community_echo_chamber_score}.

Note that our approach does not assume a specific number or size of communities and is independent of the community detection method. Moreover, it does not require prior knowledge of users' internal ideologies, setting it apart from related works~\cite{Garimella2018Quantifying,Morales2015Measuring}.


\subsection{Estimating Users' Ideology}

Our embedding model, EchoGAE, aims to position users with similar ideological leanings closer to each other in the embedding space. Therefore, we assume that we can utilize the distance in the embedding space to infer users' ideological leanings. This helps us evaluate whether EchoGAE embeds users in a way that reflects their ideology, which is the core idea behind ECS. 

After applying the EchoGAE embedding, we employ a clustering algorithm (e.g., KMeans) to detect two communities of users in the embedding space, denoted as $\omega_1$ and $\omega_2$. These communities represent the pro and anti sides of the debate, respectively. We follow similar works~\cite{Garimella2018Quantifying,Morales2015Measuring} that split the ideology spectrum into two sides.

The ideology score for each user is calculated using Equation~\ref{eq:user-idology}. It is determined by the difference between the average distance of the user $u$ to other users in $\omega_1$ and the average distance to users in $\omega_2$.

\begin{equation}
    I(u) = \frac{1}{|\omega_1|} \sum_{\substack{v \in \omega_1 \\ v \neq u}} dist(u, v) - \frac{1}{|\omega_2|}\sum_{\substack{v \in \omega_2 \\ v \neq u}} dist(u, v)
    \label{eq:user-idology}
\end{equation}

Here, $dist$ represents any distance function normalized between 0 and 1. In our implementation, we employ the Euclidean distance, but other distance measures can be used. The ideology scores $I(u)$ range from -1 to +1. Importantly, values of -1 and +1 do not inherently indicate "good" or "bad" ideologies. In Equation~\ref{eq:user-idology}, the order of the communities ($\omega_1$ and $\omega_2$) affects the sign of the ideology score. If a user belongs to $\omega_1$, their score is positive when $\omega_1$ is in the first term. Reversing the order of communities changes the sign but not the magnitude of the score. This introduces an additional layer of complexity in evaluating our method, which we address in the experimental results section.

\section{Experiments}
In this section, we present the experiments we used to assess the effectiveness of our proposed method, Echo Chamber Score (ECS), in analyzing the echo chamber effect. To evaluate its performance and reliability, we compare ECS with two commonly used methods, Random Walk Controversy (RWC) and Polarization Index (PI). Additionally, we utilize ECS to analyze echo chambers at the community level, examining the distances between users in the embedding space to gain insights into the cohesion and separation of user communities. Furthermore, We conduct an experiment to determine if the distances in the embedding space can predict the ideological leaning of users. Finally, we perform an ablation study to examine the impact of using tweets in measuring the echo chamber effect and predicting user ideology. These experiments provide valuable insights into the performance and applicability of ECS in analyzing echo chambers, predicting user ideology, and assessing the role of tweets in these measurements.

\subsection{Datasets}
To investigate the echo chamber phenomenon, we selected four topics to examine user interactions related to these subjects. Two topics were controversial: abortion and gun debates, while the other two were non-controversial: the SXSW conference and the Super Bowl. The inclusion of non-controversial topics aimed to assess our method's performance in non-polarized settings. The datasets used in our experiments are outlined in Table~\ref{tab:datasets}, and we have made them publicly available\footnote{\href{https://github.com/faalatawi/echo-chamber-score}{https://github.com/faalatawi/echo-chamber-score}} to ensure reproducibility and facilitate further research in the field of echo chamber analysis and detection.

\begin{table}
    \caption{Our datasets. We build a retweet graph for each topic where the nodes represent users, and the edges are retweet interactions. This table shows the communities of users ($\Omega$) that we detected using the Louvain Algorithm.}

    \begin{tabular}{p{0.55in}p{0.55in}p{0.55in}p{0.55in}p{0.55in}}
        \toprule
                         & Gun                                 & Abortion                                                    & Super Bowl          & SXSW                            \\
        \midrule

        \# Nodes         & 6566                                & 5087                                                        & 5460                & 2436                            \\
        \# Edges         & 14322                               & 10572                                                       & 8732                & 5325                            \\
        \midrule

        \# $\Omega$      & 2                                   & 2                                                           & 3                   & 6                               \\
        $|\Omega|$       & {[}3984, 2582{]}                    & {[}3933, 1154{]}                                            & {[}5398, 39, 23{]}  & {[}1532, 717, 85, 54, 34, 14{]} \\

        \midrule

        Keywords         & Gun, 2nd Amendment, School Shooting & Abortion Rights, Pro-Life, Pro-Choice, RoeVWade, \#LifeWins & \#SuperBowl         & SXSW                            \\
        Dates            & May-Jul 2022                        & Jun 24-25, 2022                                             & Feb 13-Mar 13, 2022 & Mar 10-19, 2022                 \\

        \midrule

        \# Labeled users & 1022                                & 1999                                                        & 194                 & 200                             \\

        \bottomrule
    \end{tabular}

    \label{tab:datasets}
\end{table}

\textbf{Data collection.} To collect data for each topic, we identified frequently used keywords in discussions (see Table~\ref{tab:datasets}) and monitored the conversation. We then gathered the retweeters of the most popular tweets associated with these keywords. This data was used to construct a graph for each topic, where users were represented as nodes, retweet interactions formed the edges, and users' tweets provided node attributes. We collected up to 200 of the users' most recent tweets (excluding retweets) to ensure an adequate amount of user-generated text for analysis. 

The gun debate dataset was collected during the period of intense debate following the Uvalde school shooting in Uvalde, Texas, on May 24, 2022. Unfortunately, school shootings in the United States often ignite polarized discussions~\cite{Guerra2013Measure} on gun violence and constitutional gun ownership rights. To capture this discourse, we selected commonly used words from both sides of the debate and monitored the conversation from May to July. We then selected the top 1200 most retweeted tweets and constructed the retweet graph. The resulting graph (shown in the lower left panel of Figure \ref{fig:embedding_and_graphs}) exhibited two communities, as identified by the Louvain algorithm \cite{Blondel2008Fast}, indicating the presence of two polarized communities~\cite{Garimella2018Quantifying}. Similarly, we collected the retweet graph from the abortion rights debate following the US Supreme Court ruling on abortion that was issued on June 24, 2022, using relevant keywords. Both the gun debate~\cite{Morini2021Standard,Guerra2013Measure,Cinelli2021Echo} and abortion~\cite{Calderon2019ContentBased,Garimella2018Quantifying} have been widely studied as topics for analyzing echo chambers and polarization.

On the other hand, for non-controversial topics, we selected the topics that have been used to study echo chambers Super Bowl~\cite{Barbera2015Tweeting} and SXSW~\cite{Garimella2018Quantifying,Emamgholizadeh2020Framework}. The Super Bowl is an annual sports event in the US, while the SXSW conference is an annual event that combines music, film, and interactive media in Austin, Texas. We followed the same data collection procedure as with the controversial topics.

\textbf{Labeling.} To evaluate the embedding quality of EchoGAE in capturing ideological similarity, we estimated users' ideological leanings. Following previous works that used news URLs to infer political leanings~\cite{Jiang2022RetweetBERT,Badawy2019Who,Bovet2019Influence,Ferrara2020Characterizing}, we obtained ideological labels for URLs from the non-partisan media watchdog AllSides\footnote{https://www.allsides.com/media-bias}. To assign labels to users, we utilized the news URLs they post as indicators of their ideology, using AllSides' political leanings for news websites' URLs. A user's political leaning is calculated as the average of the news articles they share. AllSides' ratings consist of five categories: left, center-left, center, center-right, and right, to which we assigned values of -1, -0.5, 0, 0.5, and 1, respectively. It is important to note that these values indicate opposing sides of the debate and do not inherently represent good or bad ideologies. We only used labels for users who shared at least five links. The number of labeled users for each dataset is specified in Table \ref{tab:datasets}. Notably, controversial topics tend to have more labeled users due to the nature of user engagement with these topics, as users are more likely to express their ideological leanings in these topics.

\subsection{Measuring the Echo Chamber Effect}
In this experiment, our objective is to evaluate the effectiveness of our proposed method in measuring the echo chamber effect. To accomplish this, we compare our method with commonly used techniques for calculating polarization and echo chamber effects. This comparison aims to demonstrate that our method performs comparably to existing methods and produces reliable results for measuring the echo chamber effect.

\begin{table}
\caption{Measuring the echo chamber in each dataset. ECS was able to assign a higher score for polarized topics (in red) in compassion with non-polarized ones (in blue). Please view it in color.}

\centering

    \begin{tabular}{llll}
    \hline
                                    & ECS        & RWC   & PI             \\ \hline
    \textcolor{red}{Gun}            & 0.714      & 0.419 & 0.314      \\
    \textcolor{red}{Abortion}       & 0.626      & 0.512 & 0.186      \\
    \textcolor{blue}{Super Bowl}    & 0.485      & 0.273 & 0.016      \\
    \textcolor{blue}{SXSW}          & 0.465      & 0.483 & 0.002      \\ \hline
    \end{tabular}

\label{tab:results}
\end{table}

For our experiments, we utilize two widely used baselines: Random Walk Controversy (RWC)~\cite{Garimella2018Quantifying} and Polarization Index (PI)~\cite{Morales2015Measuring}. We then compare these baselines with our proposed method, Echo Chamber Score (ECS). RWC measures the likelihood of transitioning from one community to another in a network, where a value close to one indicates polarization and close to zero indicates no polarization. On the other hand, PI measures the degree of segregation within a population by modeling the propagation of opinions based on the probability density distribution of individuals' opinions.

To compute RWC, we partition the graph into two communities using the FluidC~\cite{Pares2017Fluid} algorithm. Subsequently, we calculate the probability of transitioning from one partition to another. For PI, we employ the DeGroot opinion model~\cite{DeGroot1974Reaching} with labeled users as seeds to disseminate opinions, and then we compute the PI index for each graph. In contrast to RWC, our proposed method ECS does not require dividing the graph into two communities. The graph may consist of multiple communities, and any community detection method can be employed. In this study, we use the Louvain algorithm~\cite{Blondel2008Fast} to identify the communities, which are then used to compute ECS. Furthermore, unlike PI, our method does not rely on any labeled users, as we utilize the embeddings obtained from EchoGAE.

As shown in Table~\ref{tab:results}, our approach effectively assigns higher scores to controversial topics (e.g., Gun debate and Abortion) compared to non-controversial ones, demonstrating its ability to perform on par with existing methods. Our method aligns with PI, a highly regarded technique that employs ideology labels to gauge polarization. PI's approach closely approximates the actual labels, and our method exhibits strong agreement with it, as evidenced by a 0.99 Pearson correlation. In contrast, there are notable differences between our method and RWC. For instance, both ECS and PI indicate that the Gun Control debate is more polarized than the Abortion debate, which contradicts the findings of RWC. We posit that the requirement of RWC to partition the graph into only two communities hinders its performance. By relaxing this requirement, our measure ECS can evaluate any number of communities identified by various community detection algorithms.

These techniques (RWC, PI, and ECS) enable us to rank topics based on their polarization levels, from highest to lowest. Both PI and our method (ECS) consistently rank the topics in a similar manner. It is worth noting that our method considers the Gun debate more polarized than the Abortion debate, aligning with opinion polls. According to the Pew Research Center\footnote{https://www.pewresearch.org/}, in 2022, 61\% of Americans supported abortion access, while only 53\% advocated for stricter gun laws. This demonstrates greater disagreement and polarization within the Gun debate compared to the Abortion debate.

\subsection{Analysing the Echo Chamber Effect on Community Level}
To showcase ECS's capability in analyzing the echo chamber at a more detailed level, we conducted an experiment to examine the insights provided by our measure at the community level. The objective was to determine which community within a topic exhibited a higher level of polarization. For this experiment, we focused on the controversial topics, namely the Gun debate and Abortion, and explored how we could investigate the interaction both between and within communities. These topics were chosen due to the presence of echo chambers, as identified in the previous experiment.

Upon examining the gun dataset, we observed that the debate surrounding guns and school shootings exhibited a higher level of polarization compared to abortion, as evidenced by an ECS score of 0.714 compared to 0.626 (see Table~\ref{tab:results}). Applying the Louvain algorithm, we identified two communities in the interaction graph, with sizes of 3984 and 2582 nodes, respectively. Computing the ECS* (equation~\ref{eq:community_echo_chamber_score}) for each community, we obtained echo chamber scores of 0.739 and 0.676, indicating polarization and ideological homogeneity within both communities. Notably, the larger community demonstrated a little bit higher level of polarization.

\begin{figure}
    \centering
    \includegraphics[scale=0.45]{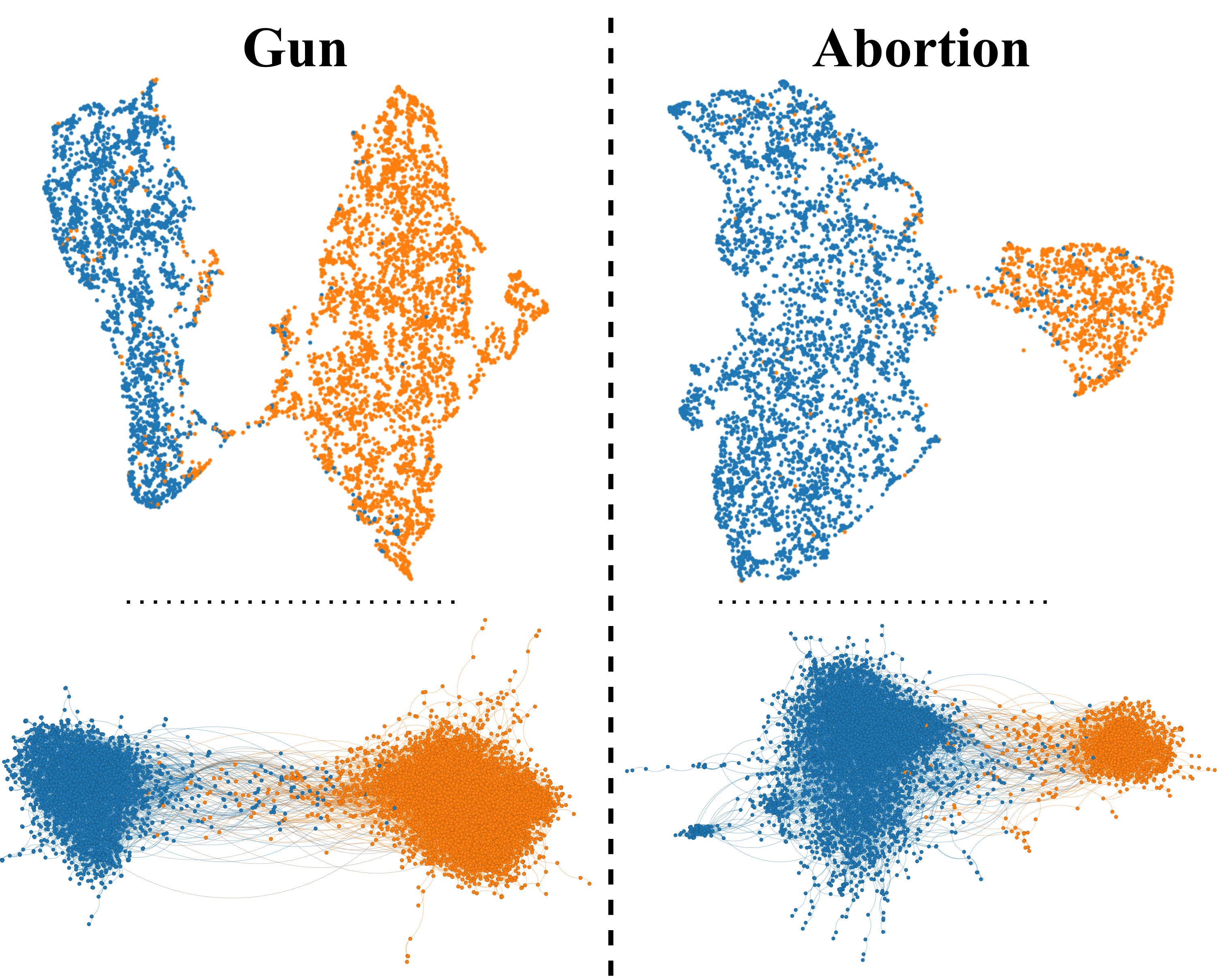}
    \caption{The top two panels show the 2D projection (using t-SNE~\cite{Maaten2008Visualizing} algorithm) of the embedding of users in each graph after embedding them using EchoGAE. The lower two panels are the two graphs. We used the ForceAtlas2~\cite{Jacomy2014ForceAtlas2} algorithm to plot the graphs. The colors represent the communities that were discovered by Louvain Algorithm}
    \label{fig:embedding_and_graphs}
\end{figure}

Upon labeling a sample of ten users from each community, we discovered that the larger community aligned with the anti-gun group, while the smaller community represented the pro-gun group. By examining the 2D projection of the EchoGAE embedding of users (refer to Figure~\ref{fig:embedding_and_graphs}), we observed that the blue community (anti-gun) appeared to have similar size as the other community (pro-gun), suggesting close levels of polarization between the two communities. However, the anti-gun higher ECS score indicates that this group is more homogenized than the other group, which is surprising. It is possible that this debate around guns is not a right and left issue, and more centrists voices are participating in the debate. This analysis would be challenging to perform using PI or RWC techniques. However, ECS, being community-independent and not reliant on ideology labels, enables such analysis without prior knowledge of community divisions and ideologies.

In the abortion dataset, we identified two communities with sizes of 3933 and 1154. The ECS* scores for these communities were 0.6 and 0.69, respectively. To gain deeper insights, we conducted a random sampling of ten users from each community and manually examined their Twitter accounts. Our analysis revealed that the larger community primarily consisted of supporters of abortion rights. On the other hand, the anti-abortion community exhibited a higher level of polarization compared to the other community. This finding aligns with the opinion polls mentioned earlier, as the anti-abortion group tends to hold a more fringe position compared to the pro-abortion group. Additionally, this alignment can be observed in the abortion rights vote that took place in Kentucky, which is considered a conservative state. During the vote, the majority of voters rejected\footnote{https://www.pbs.org/newshour/politics/kentucky-voters-reject-constitutional-amendment-on-abortion} the proposal to restrict abortion rights.

\subsection{Using Ideology Detection to Verify the Embedding Space}

We assume that the distance in the embedding space could be used to predict the political leaning of users and that users with similar ideological leanings are closer to each other in the embedding space. If we prove that the distance in the embedding space could be used to estimate the ideology of users, we could then use the distance to measure the echo chamber effect, as we rely on the distance to measure the separation (eq~\ref{eq:separation}) and cohesion (eq~\ref{eq:cohesion}) of communities in order to gauge the echo chamber effect.

 After labeling users, we then split the labeled users into training and validation sets (10\% - 90\% respectively). Since our model is unsupervised, the training set is used by the baseline model only, and we use the validation set to validate the estimation of both models. For the baseline model, we used the DeGroot opinion model~\cite{DeGroot1974Reaching}, in which the user’s ideology is the average ideology of their neighbors. After embedding users using EchoGAE, we employed the KMeans algorithm to detect two communities of users in the embedding space, referred to as $\omega_1$ and $\omega_2$, representing the pro and anti sides of the debate. Lastly, we calculated the ideology score of each user, taking into account their distances to the members of communities $\omega_1$ and $\omega_2$ in the embedding space as shown in equation~\ref{eq:user-idology}.

\begin{figure}[ht]
  \centering
  \includegraphics[scale=0.3]{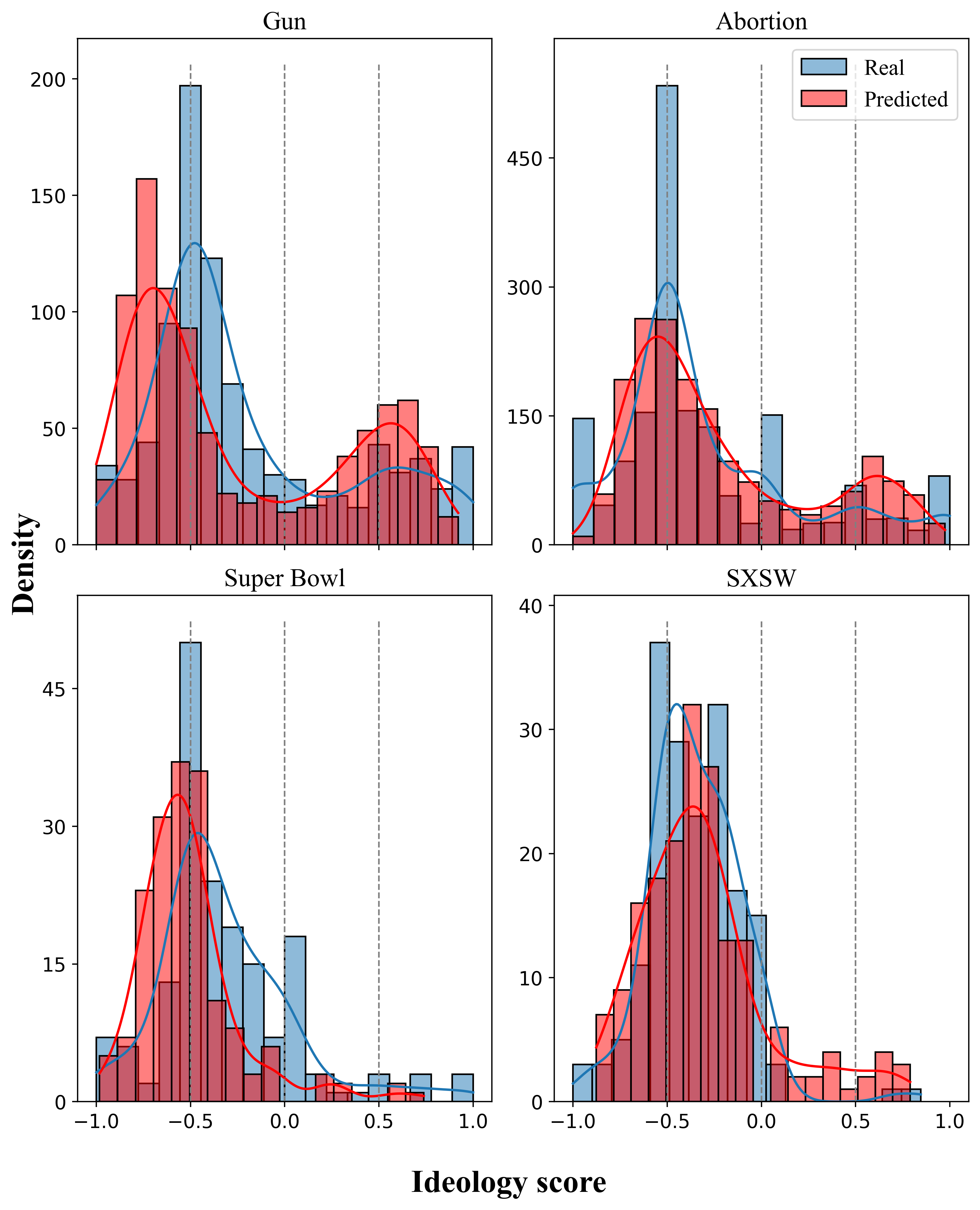}
  \caption{A histogram of the predicted vs pseudo ideology scores of users for each topic. In all the topics, the estimated (predicted) distribution of ideology scores closely matches the ideology scores estimated from the URLs that user share.}
  \label{fig:ideology_histogram}
\end{figure}

In Table~\ref{tab:ideology_results}, we present our method's outcomes for estimating ideology compared to the baseline. The resulting ideology scores were compared to the pseudo-scores obtained from AllSides labeling. Our analysis involved comparing our ideology scores to those obtained from the AllSides labeling and the baseline model using Mean Absolute Error (MAE), and Mean Squared Error (MSE). The results shown in Table \ref{tab:ideology_results} demonstrate that our model performs comparably to the semi-supervised baseline, even though our method is unsupervised (we do not use any labels in our model). Furthermore, as depicted in Figure~\ref{fig:ideology_histogram}, a high degree of concurrence is observed between the distributions of the predicted and actual ideologies.

It should be noted that in Equation~\ref{eq:user-idology}, the order of the communities (i.e., $\omega_1$ and $\omega_2$) influences the sign of the ideology score. For instance, if a user belongs to $\omega_1$ (i.e., is more closely associated with users in $\omega_1$), their ideology score would be positive if $\omega_1$ appeared in the equation's first term. However, if the order of the communities is reversed, the score's magnitude remains the same, but the sign changes. Consequently, in our measurement, we tried both orders (i.e., $\omega_1$ first then it becomes the second) and report the minimum value.


\begin{table}
\caption{We compare the predicted ideology scores from our model vs. the score from the baseline. We see that our model is comparable to the baseline in all the topics using the two measures we have.}
\centering
\begin{tabular}{@{}lllll@{}}
\toprule
           &  \multicolumn{2}{c}{MAE $(\downarrow)$}               & \multicolumn{2}{c}{MSE $(\downarrow)$} \\ \midrule
           & Our   & \multicolumn{1}{l|}{Baseline} & Our       & Baseline    \\
Gun        & 0.291 & \multicolumn{1}{l|}{0.271}    & 0.135     & 0.111       \\
Abortion   & 0.323 & \multicolumn{1}{l|}{0.269}    & 0.178     & 0.136       \\
Super Bowl & 0.352 & \multicolumn{1}{l|}{0.253}    & 0.229     & 0.147       \\
SXSW       & 0.281 & \multicolumn{1}{l|}{0.185}    & 0.148     & 0.065       \\ \bottomrule
\end{tabular}

\label{tab:ideology_results}
\end{table}

\subsection{Ablation Study}

The primary objective of this study is to examine the impact of the components of EchoGAE on the performance of two tasks: measuring the echo chamber effect and predicting the ideology of users. Specifically, the study explores the significance of using textual information, i.e., tweets, in these tasks. Table~\ref{tab:ablation_results} presents the results obtained from this study. It demonstrates that the model's performance is enhanced when tweets are utilized. This finding emphasizes the importance of linguistic similarity in measuring echo chambers and estimating ideology. Therefore, the study suggests that investing more resources to extract knowledge from tweets could lead to improved accuracy in both tasks.

However, the study also observes that good results can be achieved with the graph component alone, in situations where textual information is unavailable. Notably, even in cases where the difference in echo chamber scores between controversial and non-controversial topics is not substantial, the tweet-less model still performs well by assigning higher scores to controversial topics.

In conclusion, this study provides empirical evidence supporting the importance of incorporating textual information, such as tweets, in measuring echo chambers and estimating ideology. Nevertheless, it also highlights that satisfactory results can be obtained with graph-only models in the absence of textual data.

\begin{table}
\caption{In the ablation study, tweets (T) were used to embed users, resulting in improved performance for ECS in scoring both controversial and non-controversial topics. Additionally, using tweets for estimating users' ideology showed superior performance with lower MAE and MSE.}

\centering
\resizebox{\columnwidth}{!}{%
\begin{tabular}{lllllll}
\toprule
           & \multicolumn{2}{c}{ECS}  & \multicolumn{2}{c}{MAE $(\downarrow)$} & \multicolumn{2}{c}{MSE $(\downarrow)$} \\ 
           &  T           & No T   &  T   & No T &  T   & No T \\ \midrule
Gun        & 0.714   & 0.571   & 0.289  & 0.304  & 0.132  & 0.143  \\
Abortion  & 0.626   & 0.550   & 0.321  & 0.497  & 0.176  & 0.363  \\
Super Bowl & 0.485   & 0.438   & 0.361  & 0.428  & 0.233  & 0.315  \\
SXSW       & 0.465   & 0.464   & 0.274  & 0.421  & 0.140  & 0.286  \\ \bottomrule
\end{tabular}
}

\label{tab:ablation_results}
\end{table}

\section{Conclusion}
In this paper, we introduced Echo Chamber Score (ECS), a novel metric for quantifying echo chambers and polarization in social media networks. ECS leverages an embedding space to measure the cohesion and separation of user communities, providing insights into the echo chamber effect. To enable this measurement, we presented EchoGAE, a self-supervised user embedding model that captures ideological similarities among users and generates accurate embeddings.

Our evaluation of ECS on a Twitter dataset demonstrated its effectiveness in ranking topics based on echo chamber scores and ordering communities by polarization levels. Compared to existing metrics, ECS showcased unique capabilities in capturing the dynamics of online discourse. Our research contributes to understanding and quantifying echo chambers and polarization, which could help the development of strategies to mitigate their negative impacts and promote a more informed and open-minded society.








\bibliographystyle{IEEEtran}
\bibliography{refs}

\appendix[Deriving the ECS* equation]

Here we derive the ECS* equation. We start with the silhouette score~\cite{Rousseeuw1987Silhouettes}:
\begin{equation}
ECS^{*}(\omega) = \frac{1}{|\omega|} \sum_{u \in \omega} \left [ \frac{\Delta_u - \lambda_u}{max(\Delta_u, \lambda_u)} \right ]
\nonumber
\end{equation}

We want to scale it from 0 to 1 instead of -1 to +1. So we have:
\begin{equation}
ECS^{*}(\omega) = \frac{1}{|\omega|} \sum_{u \in \omega} \frac{1}{2} \left [ \frac{\Delta_u - \lambda_u}{max(\Delta_u, \lambda_u)} + 1 \right ]
\nonumber
\end{equation}

Multiply the terms within the square brackets by the denominator of the first fraction, $max(\Delta_u, \lambda_u)$:

\begin{equation}
ECS^{*}(\omega) = \frac{1}{|\omega|} \sum_{u \in \omega} \frac{1}{2} \left [ \frac{\Delta_u - \lambda_u + max(\Delta_u, \lambda_u)}{max(\Delta_u, \lambda_u)} \right ]
\nonumber
\end{equation}

Finally, rearrange the terms in the numerator, then separate the fractions to have the ECS equation:
\begin{equation}
ECS^{*}(\omega) = \frac{1}{|\omega|} \sum_{u \in \omega} \frac{max(\Delta_u, \lambda_u) + \Delta_u - \lambda_u}{ 2 * max(\Delta_u, \lambda_u)}
\nonumber
\end{equation}

\end{document}